      [1999/12/01 v1.4c Il Nuovo Cimento]
\documentclass{cimento}

\usepackage{graphicx}  
\title{Measurements of vector boson  plus jets at the Tevatron}
\author{L.~Cerrito \thanks{On behalf of the CDF and D0 Collaborations}
}
\institute{Department of Physics, Queen Mary University of London -- London (UK)}

\PACSes{
\PACSit{12.38.Qk}{Experimental tests of QCD}
\PACSit{14.70.--e}{Gauge bosons}
\PACSit{14.65.--q}{Quarks}}

\begin{document}

\maketitle
\vspace{-3cm}
\begin{abstract}
We present preliminary measurements of $Z/\gamma^*$+jets, $W+c$ and $Z+b+X$ at the Tevatron, and review recent measurements of vector boson plus inclusive and heavy-flavor jets production. All measurements are in agreement with next-to-leading-order QCD calculations within the experimental and theoretical uncertainties. We also point to comparisons of the production rate and kinematics of the data with several Monte Carlo simulation programs of vector boson + jets processes.
\end{abstract}

\section{Introduction}
The associated production of vector boson (V) and jets is a major background to single and pair-produced top quarks, as $W$ and $Z+$ heavy flavor jets share much of the final-state signature of top events. The contribution of $W$+ heavy flavor quark ($c,~b$) for instance leads to one of the main systematic uncertainties to the identification and measurements of single top at the Tevatron, but also searches for the Higgs boson in the $WH\rightarrow Wb\bar{b}$ channel. In recent measurements of top quarks at CDF \cite{ttbarxsec}, the estimates of the $Wb\bar{b}$, $Wc\bar{c}$ and $Wc$ backgrounds carry a precision of only $\sim\pm30$\%. Calculations of V+jets at next-to-leading-order (NLO) in QCD are available mostly for lower jet multiplicities. Moreover, a number of Monte Carlo (MC) simulation tools are also available: these include simulations based on leading-order (LO) matrix elements followed by parton shower modeling (e.g. {\tt PYTHIA}, {\tt HERWIG}), or higher-order matrix elements for the hard-scattering process (e.g. {\tt ALPGEN}, {\tt SHERPA}), followed by modeling of the parton shower. The production of V+jet, in addition to be crucial for the understanding of top quark candidate samples, gives important information on parton distribution functions (PDFs) inside the proton and allows for stringent QCD and Standard Model tests. We present preliminary measurements of $Z/\gamma^*$+jets, $W+c$ and $Z+b+X$ and review the latest measurements of V + inclusive and heavy flavor jets at the Tevatron.


\section{\boldmath Vector boson plus inclusive jets}
\subsection{\boldmath \bf {$W$+jets}}
The production cross section of $W+$ jets in $p\bar{p}$ collisions at $\sqrt{s}=1.96$~TeV has been measured \cite{wjets} using a data sample corresponding to an integrated luminosity of $\sim$320 pb$^{-1}$, collected with the CDF~II detector. The CDF~II detector is described in detail elsewhere \cite{cdfdet1,cdfdet2,cdfdet3,cdfdet4}. $W$ bosons are identified in their electron decay channel, and electrons from the $W$ decay are required to have transverse energy $E_T>20$~GeV and pseudo-rapidity $|\eta|<1.1$ \cite{eta}. Jets are reconstructed using a cone algorithm \cite{cone} with radius 0.4 and are required to have $E_T^{jet}>20$~GeV and $|\eta|<2.0$, where $E_T^{jet}$ is calibrated such that it measures on average the summed $E_T$ of the particles within the jet cone that are the result of the $p\bar{p}\rightarrow W+X$ interaction. The measurement's results include the determination of the total cross section: $\sigma(p\bar{p}\rightarrow W+\geq n-{\rm jet};E_T^{nth-jet}>25~{\rm GeV})\times BR(W\rightarrow e\nu)$ and the differential cross section $d\sigma(p\bar{p}\rightarrow W+\geq n-{\rm jet})/dE_T^{n{\rm-th jet}}\times BR(W\rightarrow e\nu)$ for ($n=1-4$). The data is seen to agree to better than $\sim 10$\% with NLO calculations done with the {\tt MCFM} \cite{mcfm} program ($n\leq 2$). The analysis also presents comparisons of the data with MC modeling based on LO perturbative QCD (pQCD) calculation plus parton shower and hadronization ({\tt MADGRAPH}+{\tt {PYTHIA}} \cite{smpr} and {\tt ALPGEN}+{\tt HERWIG} \cite{mlm}). The production cross section measured in the data ($\sigma_n$), as a function of $n$-jets, results higher than that predicted by these models by about 40\%, although the ratio $\sigma_n/\sigma_{n-1}$ is well reproduced.

\subsection{\boldmath \bf{$Z/\gamma^*$+jets}}
\begin{figure}[h]
\begin{center}
\begin{tabular}{cc}
\includegraphics[width=6.7cm]{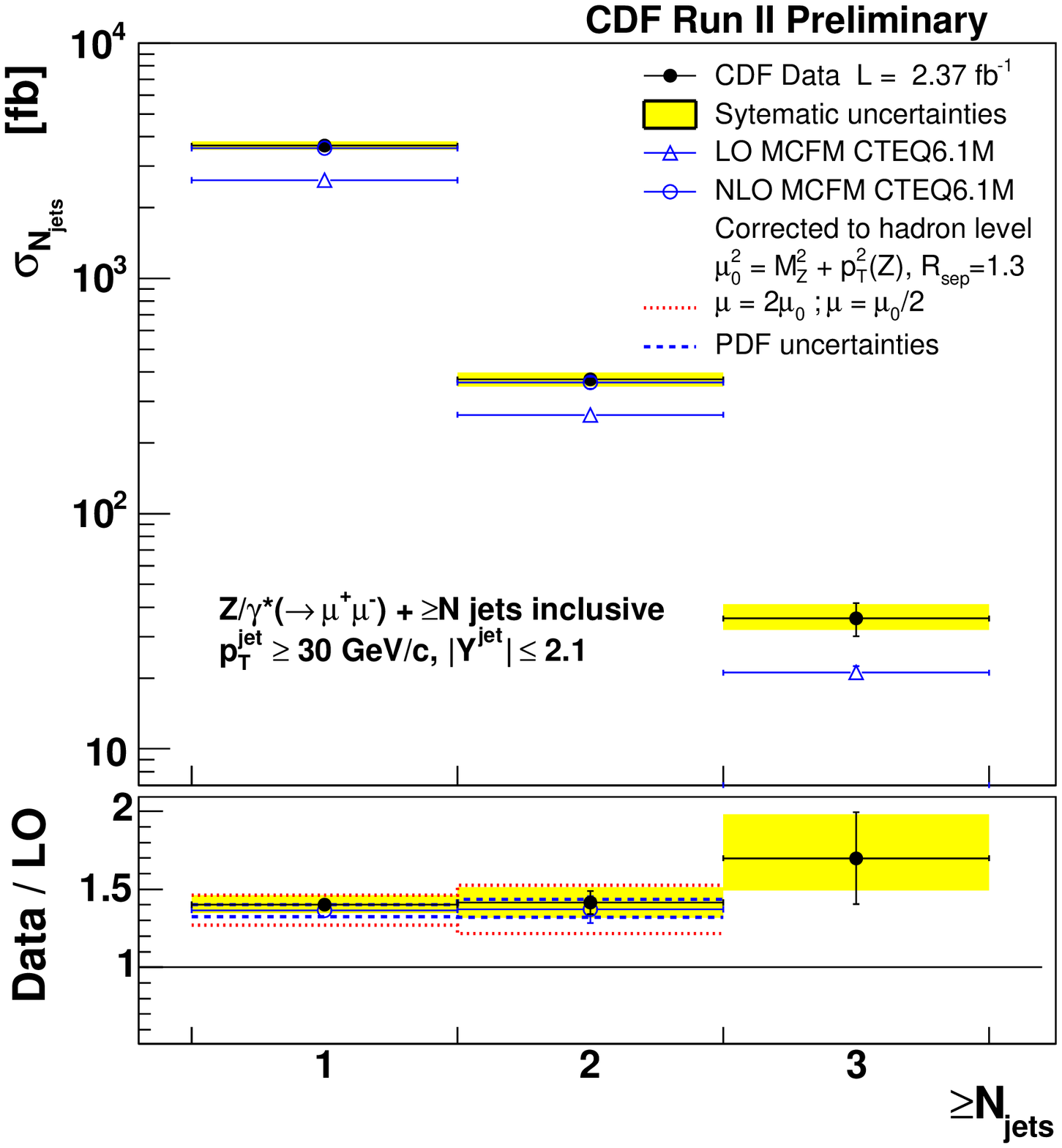}   & 
\includegraphics[width=6.7cm]{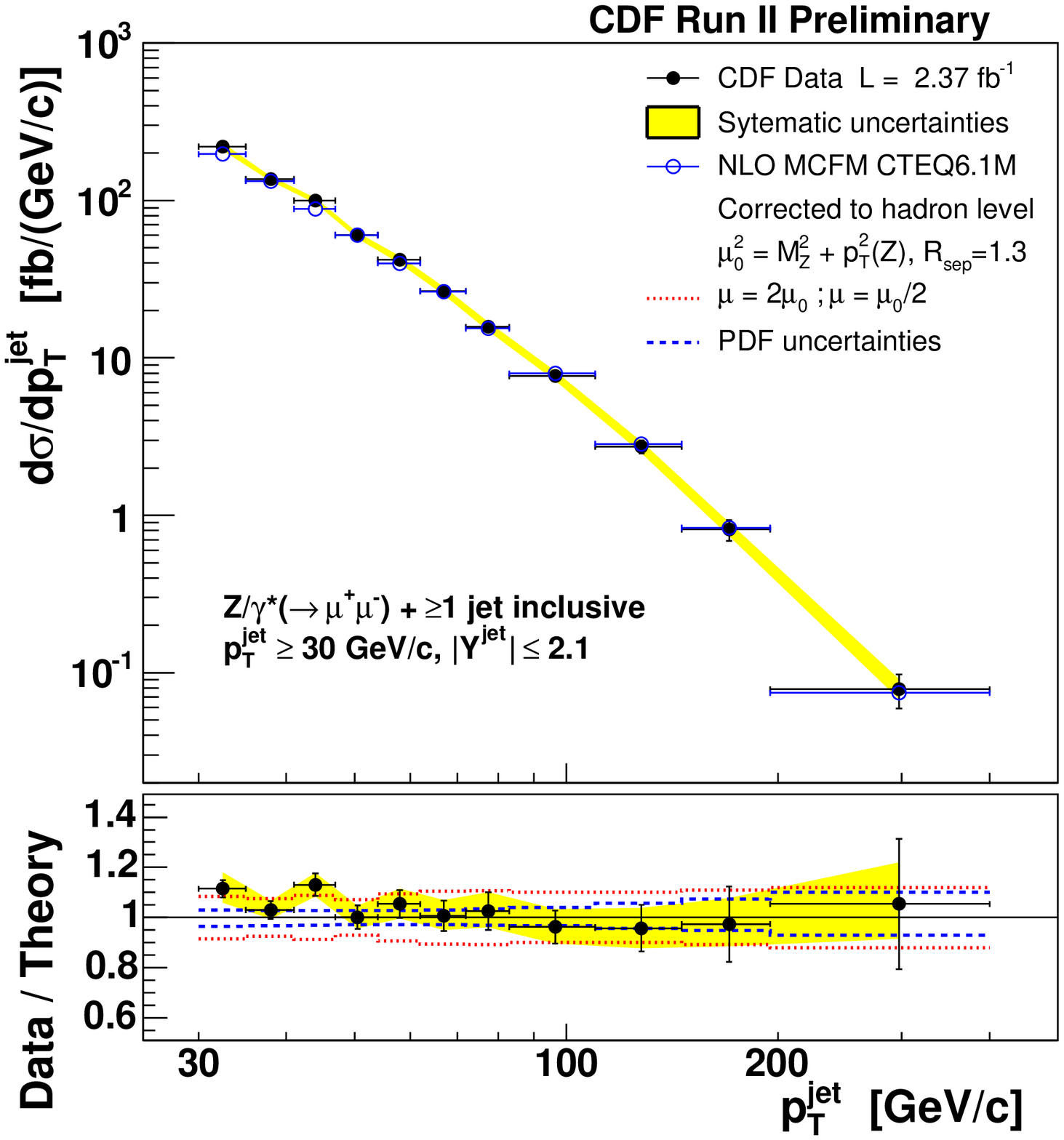}   \\
(a) & (b) \\
\end{tabular}
\caption{Measured total cross section for inclusive jet production in $Z/\gamma^*\rightarrow \mu\mu$ events, as a function of the number of jets (a) and the jet $p_T$ (b). The data are compared to LO and NLO pQCD predictions.}
\label{fig:zjets}
\end{center}
\end{figure}

The production of $Z/\gamma^*$+jets at the Tevatron has been measured in the $Z\rightarrow ee$ decay channel \cite{zee_cdf} by the CDF collaboration and in $Z/\gamma^*\rightarrow\mu\mu$ and $Z/\gamma^*\rightarrow ee$  decay channels \cite{zmm_d0,zmm_d0_2,zee_d0} by the D0 collaboration, using $\sim$2.5~fb$^{-1}$ and $\sim$1.0~fb$^{-1}$ of collision data respectively. We present here a preliminary measurement in the $Z/\gamma^*\rightarrow\mu\mu$ decay channel using $\sim$2.4 fb$^{-1}$ of data collected with the CDF~II detector. Jets are reconstructed using the midpoint algorithm \cite{midpoint} with cone radius $R=0.7$ and are required to have transverse momentum $p_T^{jet}\geq 30$~GeV/$c$, rapidity $|y^{jet}|\leq 2.1$ \cite{rapidity} and a minimum distance between the jets and the muons $\Delta R_{\mu-jet}>0.7$. The value of $p_T^{jet}$ is corrected back to the particles that enter the jet cone, and includes also corrections for the multiple $p\bar{p}$ interactions per crossing at high instantaneous luminosity. In Fig \ref{fig:zjets} the data is compared to LO and NLO predictions computed using the {\tt MCFM} program. Fig.~\ref{fig:zjets}(a) shows the cross section as a function of the number ($N_{jets}$) of jets in the event, while Fig.~\ref{fig:zjets}(b) shows the inclusive $p_T^{jet}$ differential cross section for $Z/\gamma^*\geq 1$ jet. The theoretical predictions include parton-to-particle level correction factors that account for underlying event and fragmentation processes. The measurements are well described by NLO pQCD predictions (for $N_{jet}\geq3$ only LO predictions are available), and the $N_{jet}$ differential cross section measurement suggests a common LO-to-NLO multiplicative factor of $\sim 1.4$, approximately independent of $N_{jet}$.

\begin{figure}[ht]
\begin{center}
\includegraphics[width=12.5cm]{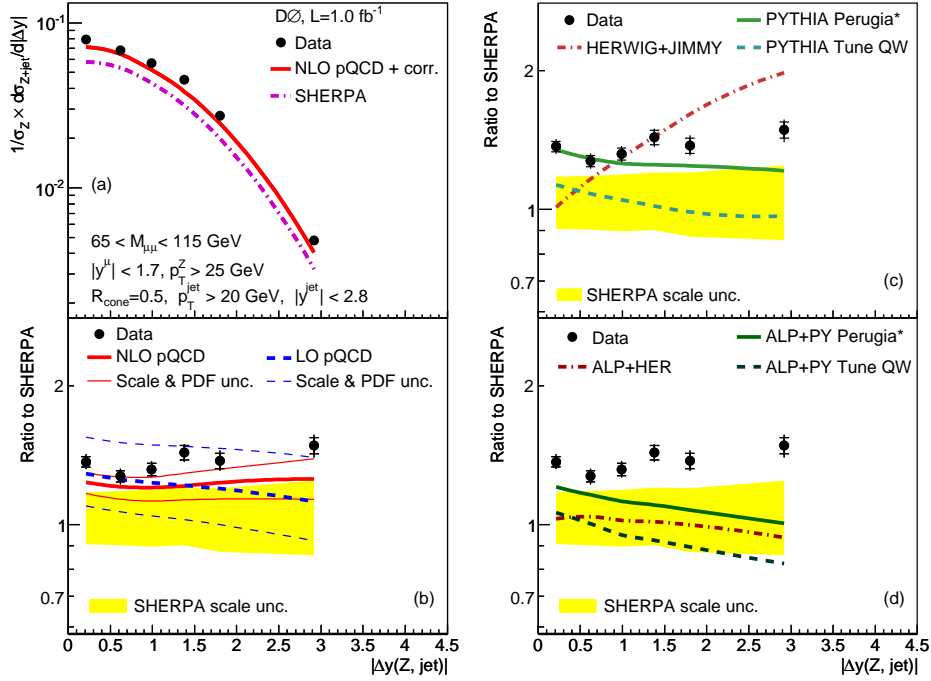}  
\caption{Measured (normalised) cross section for $Z/\gamma^*+{\rm jet}+X$ production  with $p_T^Z>25$~GeV/$c$, in bins of $|\Delta y(Z,{\rm jet})|$ \cite{zmm_d0_2}. All ratios in (b), (c) and (d) are shown relative to the prediction from {\tt SHERPA}.}
\label{fig:d0_mm_2}
\end{center}
\end{figure}

The extensive $Z$-boson+jets sample collected at the Tevatron has also allowed detailed comparisons of the data with LO and NLO predictions in a variety of phase space regions. In the most recent of these analyses \cite{zmm_d0_2}, differential $Z/\gamma^*$+jet+X cross sections are measured, binned in the azimuthal angle between the $Z/\gamma^*$ and leading jet, $\Delta\phi(Z,jet)$, absolute value of the rapidity difference between the $Z/\gamma^*$ and leading jet, $|\Delta y(Z,jet)|$, and the absolute value of the average rapidity of the $Z/\gamma^*$ and leading jet, $|y_{\rm boost}(Z+{\rm jet})|$. The $\Delta\phi(Z,{\rm jet})$ distribution is sensitive to QCD radiation. The rapidity variables, $\Delta y(Z,{\rm jet})$ and $y_{\rm boost}(Z+jet)$ have primary contributions from the relative momenta of the incoming partons in the hard scatter but are modified by any additional QCD radiation. Fig. \ref{fig:d0_mm_2} shows the normalized differential cross section as a function of $|\Delta y(Z,{\rm jet})|$. The data corresponds to an integrated luminosity of 0.97$\pm$0.06 fb$^{-1}$ collected with the D0 detector. A full description of the D0 detector is available elsewhere \cite{d0detector}, jets are reconstructed using an iterative midpoint algorithm \cite{d0_jetalgo} with cone radius 0.5 and their transverse momentum is corrected for the calorimeter response, instrumental noise and pile-up from multiple $p\bar{p}$ interactions and bunch crossings. The analysis selects events with $p^{jet}_T>20$~GeV/$c$ and $|\eta^{jet}|<2.8$. The data are compared with LO and NLO pQCD calculations obtained using {\tt MCFM}, after correcting the predictions for hadronization and underlying event using {\tt PYTHIA}. The data are also compared in Fig.~\ref{fig:d0_mm_2}(c)(d) with matrix element calculations with matched parton shower implementations such as {\tt SHERPA} and {\tt ALPGEN}. The NLO pQCD calculation provides a good  modeling of the data and is a significant improvement in both shape and uncertainty over LO.

\subsection{\boldmath \bf{$\gamma+$jets}}
The production of $p\bar{p}\rightarrow\gamma+{\rm jet}+X$ at a center-of-mass energy $\sqrt{s}=1.96$~TeV has been measured \cite{d0_phojet} using $\sim$1.0~fb$^{-1}$ of data collected by the D0 detector. Photons are reconstructed in the central rapidity region $|y^\gamma|<1.0$  with transverse momenta in the range $30<p_T^\gamma<400$~GeV/$c$ while jets are reconstructed in either the central $|y^{jet}|<0.8$ or forward $1.5<|y^{jet}|<2.5$ rapidity intervals with $p_T^{jet}>15$~GeV/$c$. Different angular configurations between the photon and the jets can be used to simultaneously test the dynamics of QCD hard-scattering subprocesses in different regions of parton momentum fraction $x$ and large hard-scattering scales $Q^2$. The data are compared to NLO QCD predictions obtained using {\tt JETPHOX} \cite{jetphox1,jetphox2,jetphox3} with {\tt CTEQ6.5M} PDF \cite{cteq6.5} and {\tt BFG} fragmentation functions of partons to photons \cite{bfg}. While the NLO predictions are seen to give a good overall description of the $d\sigma/dp_T^\gamma$ differential cross section, discrepancies of order 30--40\% appear when attempting the describe simultaneously the data in four separate kinematic regions defined by the rapidity of the photon and that of the leading jet (central jets with $y^\gamma\cdot y^{jet}>0$ and $y^\gamma\cdot y^{jet}<0$, and forward jets with $y^\gamma\cdot y^{jet}>0$ and $y^\gamma\cdot y^{jet}<0$). The measurements are compared to the corresponding theoretical predictions, which are seen to overestimate slightly the production measured for photons with $p_T^\gamma>200$~GeV/$c$.

\section{Vector boson plus heavy-flavor jets}
\subsection{\boldmath \bf{$W+b+X$}}
The measurement of the production of $W+b$-jets provides and important test of QCD, and the understanding and modeling of this process is crucial e.g. in searches for standard model Higgs boson via $WH\rightarrow Wb\bar{b}$ and measurements of single top. Theoretical predictions for $W+b$-jets production are based on NLO calculations in the 1-jet and 2-jet multiplicities \cite{wbbtheo1,wbbtheo2,wbbtheo3} and show an enhancement over LO of up to a factor of two. The cross section for jets from $b$ quarks produced with a $W$ boson has been measured \cite{wbb} in $p\bar{p}$ collision data at $\sqrt{s}=1.96$~TeV using $\sim$1.9~fb$^{-1}$ of integrated luminosity recorded with the CDF~II detector. The measurement is based on the selection of events consistent with the electronic or muonic decay of a $W$ boson and containing one or two jets. The identification of heavy flavor-originated jets is based on the detection of a secondary vertex (vertex $b$-tagging), reconstructed from the charged particles within each jet, and well displaced from the primary $p\bar{p}$ interaction vertex location. The invariant mass of the charged particles forming the secondary decay vertex is used as a discriminant to separate $b$-jets from charm and certain light flavor hadrons which are accidentally $b$-tagged. The measurement's result: $\sigma_{WbX}(p_T^{e,\mu}>20~{\rm GeV}/c,~|\eta_{e,\mu}|<1.1,~p_{T,\nu}>25~{\rm GeV}/c,~E_{T,bjet}>20~{\rm GeV},~|\eta_{bjet}|<2.0)\times BR(W\rightarrow\ell\nu)=2.47\pm0.27({\rm stat})\pm0.42({\rm syst})~{\rm pb}$ is in agreement, although about two standard deviations higher, with the NLO calculation for the corresponding phase space of 1.22$\pm$0.14~pb. Modelings based on LO from {\tt PYTHIA} and summed fixed-order from {\tt ALPGEN} of the jet $E_T$ and jet $\eta$ differential cross sections are seen to agree with the data, although the overall production cross section predicted by these programs is lower than the measured value by factors of 2.5--3.5.

\subsection{\boldmath \bf{$W+c$}}
Calculations of the production of $W+$charm quark at the Tevatron carry a precision of only $\sim$30\%, due to uncertainties in the strange-quark density in the proton and on the renormalization and factorization scales in the NLO calculation \cite{wcharm_theo}. Therefore, experimental constraints are crucial to help in searches and measurement of signatures with the $W$+heavy flavor final state. The experimental determination of $W$+charm is based on  the correlation between the charge of the $W$ boson and the charge of the lepton ($e$ or $\mu$) from the semileptonic decay of the charm hadron. Charge conservation in the process $gq\rightarrow Wc$ ($q=d,s$) allows as final states only the pairings $W^+\bar{c}$ and $W^-c$ and therefore the charge of the lepton from the semileptonic decay of the $c$ and the charge of the lepton from the $W$ decay are always of opposite sign. By subtracting the number of same-sign combinations from the opposite-sign sample it is possible to highlight a $W$+charm signal from which the production cross section is measured. We report a preliminary measurement using the semileptonic electron decay of the charm, performed with $\sim$4.3~fb$^{-1}$ of collision data recorded by the CDF~II detector. The analysis's result: $\sigma_{W+c}(p_{T,c}>20~{\rm GeV}/c,~|\eta_c|<1.5)\times BR(W\rightarrow\ell\nu)=21.1\pm7.1({\rm stat})\pm 4.6({\rm syst})~{\rm pb}$ is in agreement with the NLO theoretical calculation of $11.0^{+1.4}_{-3.0}~{\rm pb}$. More precise measurements of $W+c$ at the Tevatron have been derived using the semileptonic muon decay of the charm, as the identification of soft muons in jets leads to purer charm samples than those based on jets with soft electrons. The two most accurate measurements to-date of $W$ boson production with a single charm quark in $p\bar{p}$ collisions at $\sqrt{s}=1.96$~TeV, have been made using $\sim$1.8~fb$^{-1}$ and $\sim$1.0~fb$^{-1}$ of data collected with the CDF~II and D0 detectors respectively \cite{wcharm_cdf,wcharm_d0}. The CDF anaysis measures $\sigma_{Wc}(p_{Tc}>20~{\rm GeV}/c,|\eta_c|<1.5)\times {\rm BR}(W\rightarrow\ell\nu)=9.8\pm3.2~{\rm pb}$. The D0 analysis measures the ratio $\sigma_{W+c}/\sigma_{W+jets}=0.074\pm0.019({\rm stat})^{+0.012}_{-0.014}({\rm syst})$, in agreement with a LO prediction of 0.044$\pm$0.003. Both these measurements are based on the identification of the charm-jet via the semileptonic muon decay of the charm. A precision of $\sim\pm15$\% could be reached in the near future by analyzing the larger datasets of several fb$^{-1}$ already available at the Tevatron.

\subsection{\boldmath \bf{$Z+b+X$}}
\begin{figure}[ht]
\begin{center}
\includegraphics[width=6.5cm]{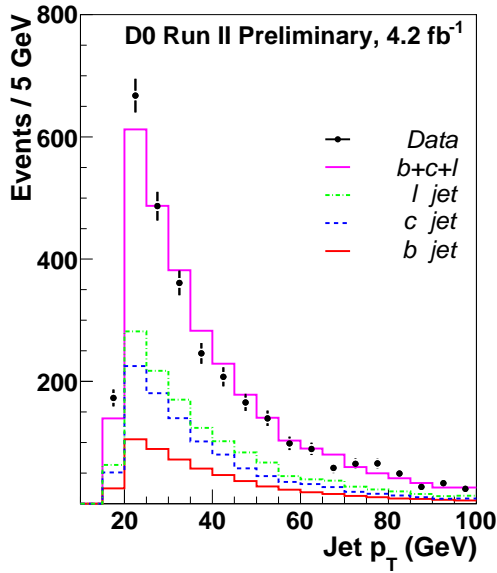}  
\caption{The leading jet $p_T$ in $Z+b$-jet candidate events, with contributions from $b$, $c$ and mis-identification of light-flavor jets.}
\label{fig:zbbpt}
\end{center}
\end{figure}

Similarly to $W+b$ or $c$ quarks, the associated production of $Z$ with one or more $b$ jets provides an important test of QCD and  its accurate modeling is crucial in particle searches, e.g. for the standard model Higgs boson in the $ZH\rightarrow Zb\bar{b}$ decay mode \cite{zbb_higgs}. We present a preliminary measurement of $Z+b$ jets performed using $\sim$4.2~fb$^{-1}$ of data collected with the D0 detector. Jets are reconstructed using the iterative midpoint algorithm \cite{d0_jetalgo} with a cone of radius $\Delta R=0.5$. The energy of jets is corrected for detector response, noise, the presence of additional interactions, and energy deposited outside the reconstructed jet cone. Figure \ref{fig:zbbpt} shows the $p_T$ distribution of the candidate $b$ jets in data and MC, where the MC expectations for various jet flavors have been weighted by the fractions determined by a maximum likelihood fit of the distribution of probability that a jet originates from the primary vertex. This analysis supersedes an earlier measurement \cite{zb_d0} and yields: $\sigma(Z+b~jet)/\sigma(Z+jet)=0.0176\pm0.0024({\rm stat})\pm0.0023({\rm syst})$ for jets with  $p_T>20~{\rm GeV}/c$ and $|\eta|\leq 1$. The result is compatible with the NLO QCD prediction of 1.8$\pm$0.4\% \cite{zb_theo}. A measurement of $Z+b+X$ had also been performed using $\sim$2~fb$^{-1}$ of collision data collected by the CDF~II detector \cite{zb_cdf}. In the analysis, jets are selected with $E_T>20~{\rm GeV}$ and  $|\eta|<1.5$ and are identified as $b$ jets by the detection of a secondary vertex. The measurement determines the ratio of the integrated $Z+b$ jet cross section to the inclusive $Z$ production cross section: (3.32$\pm$0.53(stat)$\pm$0.42(syst))$\cdot 10^{-3}$. The production ratio $\sigma(Z+b~jet)/\sigma(Z+jet)$ for jets and $b$ jets with $E_T>20$~GeV and $|\eta|<1.5$ is measured as $(2.08\pm0.33({\rm stat})\pm0.34({\rm syst}))$\%, both values consistent with predictions from LO MC generators ({\tt ALPGEN} and {\tt PYTHIA}) and NLO QCD calculations using {\tt MCFM}.

\subsection{\boldmath \bf {$\gamma+b+X,~\gamma+c+X$}}
The production of $\gamma$+heavy flavor jets has been measured \cite{d0_gammahf} using an integrated luminosity of $\sim$1 fb$^{-1}$ of data collected by the D0 detector. Photons produced in association with heavy quarks ($c$ or $b$) provide valuable information about the parton ($b,~c$ and gluon) distributions in the proton and anti-proton. Measurements of the dynamics of $\gamma+b,c+X$ also provides testing of perturbative QCD and its modeling helps in searches for new physics that shares the $\gamma$+heavy flavor final states. The analysis provides the first measurements of the differential cross sections $d^2\sigma/(dp_T^\gamma dy^\gamma dy^{jet})$ in $\gamma+b,c$ candidate events, for photons with transverse momenta $30<p_T^\gamma<150$~GeV/$c$, photon rapidities $|y^\gamma|<1.0$, jet rapidities $|y^{jet}|<0.8$, and jet transverse momenta $p_T^{jet}>15$~GeV/$c$. The $\gamma+b+X$ production is seen to agree with the prediction based on NLO pQCD using {\tt CTEQ6.6M} PDFs over the entire range of $p_T^\gamma$. The $\gamma+c+X$ production is seen to agree with the NLO pQCD prediction for $p_T^\gamma< 50$~GeV/$c$ whereas the data becomes larger by about 2 standard deviations than the calculated cross section for harder photons ($p_T^\gamma>70$~GeV/$c$).

\section{Conclusions}
We have presented preliminary measurements of $Z/\gamma^*$+jets, $W+c$ and $Z+b+X$, and reviewed the latest measurements of vector boson plus inclusive and heavy flavor jets at the Tevatron. All measurements of the production and kinematics of V+jets are in good agreement with NLO pQCD calculations. More precise measurements of $W+b+X$ and $W+c$ are needed to constrain the theoretical uncertainties, and help reducing uncertainties on measurements of single top quark and $WH\rightarrow b\bar{b}$ searches. Significant improvements on measurements of $W+b,c$ production are expected in the near future based on the latest Tevatron datasets and developments of the analyses technique.

\acknowledgments


\begin{thebibliography}{0}
\bibitem{eta} We use a ($z$, $\phi$, $\theta$) coordinate system where the $z$ axis is in the direction of the proton beam, and $\phi$ and $\theta$ are the azimuthal and polar angles, respectively. The pseudorapidity is $\eta=-ln({\rm tan}\frac{\theta}{2})$. Transverse energy and momentum are $E_T=E{\rm sin}\theta$ and $p_T=p{\rm sin}\theta$, respectively, where $E$ and $p$ are energy and momentum.
\bibitem{rapidity} Rapidity is defined as $y=0.5 {\rm ln}[(E+p_z)/(E-p_z)]$, where $E$ is the energy, and $p_z$ is the component of momentum parallel to the proton beam direction.
\end{thebibliography}
\end{document}